\documentclass[12pt]{iopart}

\usepackage{iopams,lineno,bm,multirow,adjustbox}  
\usepackage{graphicx, caption}
\usepackage{url,hyperref}
\DeclareMathSymbol{\mh}{\mathord}{operators}{`\-}
\begin{document}

\title[Transient soft X-ray absorption in solids from TDDFT]{Simulation of attosecond transient soft X-ray absorption in solids using generalized Kohn-Sham real-time TDDFT}

\author{C. D. Pemmaraju}

\address{Stanford Institute for Materials and Energy Sciences, SLAC National Accelerator Laboratory, Menlo Park, CA 94025, USA}
\ead{dasc@slac.stanford.edu}
\vspace{10pt}
\date{\today}
\begin{abstract}
Time-dependent density functional theory (TDDFT) simulations of transient core-level spectroscopies require a balanced treatment of both valence- and core-electron excitations. To this end, tuned range-separated hybrid exchange-correlation functionals within the generalized Kohn-Sham scheme offer a computationally efficient means of simultaneously improving the accuracy of  valence and core  excitation energies in TDDFT by  mitigating delocalization errors across multiple length-scales. In this work range-separated hybrid functionals are employed in conjunction with the velocity-gauge formulation of real-time TDDFT to simulate static as well as transient soft X-ray near-edge absorption spectra in a prototypical solid-state system, monolayer hexagonal boron nitride, where excitonic effects are important. In the static case, computed soft X-ray absorption edge energies and line shapes are seen to be in good agreement with experiment. Following laser excitation by a pump pulse,  soft X-ray probe spectra are shown to exhibit characteristic features of population induced bleaching and transient energy shifts of exciton peaks. The methods outlined in this work therefore illustrate a practical means for simulating attosecond time-resolved core-level spectra in solids within a TDDFT framework.
\end{abstract}

%
\vspace{2pc}
\maketitle
%
%

\section{Introduction}\label{intro}
Over the past twenty years, rapid advances have been realized in the ability to generate extremely short X-ray laser pulses in a variety of settings ranging from table-top high-harmonic generation (HHG)~\cite{Paul2001,Hentschel2001} setups to large-scale synchrotron and X-ray free-electron laser (XFEL)~\cite{Emma2010,Duris2020} facilities.  This in turn has led to the development of a wide variety of new time-domain X-ray spectroscopic techniques capable of investigating electron and ion dynamics in matter with picosecond to attosecond time-resolution~\cite{Leone2014,Ramasesha2016,Kraus2018a}. Due to their characteristic element-specificity and unprecedented temporal resolution, ultrafast extreme-UV (XUV) and X-ray spectroscopies are emerging as indispensable tools in chemistry and materials science~\cite{Marangos2019,Kraus2018a}.  In particular, X-ray transient absorption spectroscopy (XTAS) which deploys short X-ray pulses to probe ultrafast modulations in the X-ray absorption near-edge structure (XANES) is now increasingly employed to investigate electron dynamics in molecular and solid-state systems~\cite{Ramasesha2016,Kraus2018a,Geneaux2019}.  
In the case of static X-ray absorption spectroscopy (XAS)~\cite{Groot2008} which is widely used for characterizing the ground state electronic structure of materials, the interpretation of experimentally measured spectra has been greatly facilitated over the years by the development of theoretical tools for atomistic simulation of XANES using quantum mechanical methods~\cite{Rehr2009}.  A wide variety of first-principles approaches now exist for static XANES simulations and have been reviewed in the literature~\cite{Groot2008,Rehr2009,Prendergast2006,Petrenko2008,Besley2010,Gilmore2015,Maganas2019,Pinjari2014,Zhang2012a}. The advent of XTAS similarly requires the concomitant development of theoretical techniques and tools to aid the interpretation of time-resolved X-ray spectra that probe the dynamics of quantum mechanical excited states of matter~\cite{Geneaux2019}. In this context, the framework of time-dependent density functional theory (TDDFT)~\cite{Runge1984,Marques2012a,Ullrich2012} which has long been a workhorse method for excited-state simulations in quantum chemistry represents an attractive option for its potential to balance computational feasibility and accuracy. For simulating the response of materials to  intense ultra-short laser pulses relevant to pump-probe experiments in particular, real-time TDDFT (RT-TDDFT)~\cite{Marques2012a,Ullrich2012,Yabana1996} offers a convenient formalism. Hence in recent years, RT-TDDFT has been deployed for modeling femto- to attosecond timescale transient absorption in both molecules and solids for energies  in the infrared (IR) to XUV range~\cite{DeGiovannini2013,Fischer2015a,Nguyen2016,Otobe2016,Sato2018}.  To date however, RT-TDDFT simulations of XTAS involving core-excitations in the soft X-ray energy range have not been reported especially in extended systems.  In this work therefore, the velocity-gauge formulation of RT-TDDFT (VG-RT-TDDFT)~\cite{Bertsch2000,Yabana2012,Pemmaraju2017,Pemmaraju2018b} is employed to simulate XTAS in monolayer hexagonal boron nitride (h-BN) which is a prototypical 2D-periodic insulator characterized by strong excitonic effects~\cite{Ferreira2019,Pemmaraju2018b}. 

XTAS pump-probe experiments~\cite{Geneaux2019,Kraus2018a} typically involve excitation by photon energies on the order of $\sim$1 eV from intense IR-UV $pump$ laser pulses to drive valence electron dynamics in a sample while  XUV/X-ray $probe$ pulses investigate the pump-induced excited-state dynamics of valence electrons by recording time-dependent modulations of core-electron excitation features in the $\sim$10-1000 eV range. Therefore, a balanced theoretical description of XTAS needs among other things, to provide a reliable treatment of both valence and core electronic excitations simultaneously. This necessarily translates in a TDDFT context, to a need for choosing appropriate exchange-correlation (XC) potentials~\cite{Marques2012a,Ullrich2012} which primarily dictate the accuracy of the simulations. As noted by Zhang et al~\cite{Zhang2015}, similar considerations apply in emerging nonlinear X-ray spectroscopy protocols where nonlinear excitation of core-electrons by intense coherent X-ray pulses can be used to in turn drive valence electron dynamics.  While the choice of appropriate XC functionals is generally an open problem in TDDFT, several useful guidelines have been proposed in the literature over the past two decades~\cite{Dreuw2004,Petrenko2008,Song2008,Besley2010,Kronik2012,Imamura2015,Refaely-Abramson2015a,Yang2015,Byun2020}. Firstly, in the case of valence excitations, local, semi-local and short-range nonlocal hybrid functionals  even where reliable for ground state total energy calculations, are prone to significant errors in their TDDFT description of charge-transfer excitations in molecular systems~\cite{Dreuw2004,Song2008,Kronik2012,Imamura2015} and excitonic effects in semiconducting or insulating solids~\cite{Refaely-Abramson2015a,Yang2015}. To mitigate these shortcomings, long-range corrected (LRC) XC functionals~\cite{Dreuw2004,Song2008,Kronik2012,Refaely-Abramson2015a,Yang2015} within the \textit{generalized} Kohn-Sham (GKS) approach~\cite{Seidl1996} have been proposed and applied successfully in many molecular and solid-state systems. However, LRC-XC functionals that provide a reliable description of valence electronic excitations by improving the long-range asymptotic behavior of the XC potential~\cite{Refaely-Abramson2015a,Yang2015}  do not necessarily improve the description of localized core-electrons and their excitations relevant to XUV and X-ray spectroscopies~\cite{Besley2010, Imamura2015,Pemmaraju2018b}. In particular, as shown recently~\cite{Pemmaraju2018b}, LRC-XC functionals can significantly underestimate core-excitonic effects in extended systems. In the context of molecular systems, a number of groups have  proposed short-range corrected (SRC) functional forms that are either tailored specifically to improve the description of core-excitations~\cite{Besley2010} or more generally to provide a balanced treatment of both core and valence electronic excitations simultaneously~\cite{Song2008,Imamura2015}.  In this work a simple procedure is proposed to incorporate short-range corrections for core-electrons in an extended periodic system such as 2D h-BN. It is shown that by augmenting a screened LRC-XC functional with a SRC component, the description of core-level excitonic effects is significantly improved providing a balanced treatment of valence and core-excitations in VG-RT-TDDFT.  

\section{Theoretical approach}\label{theory}
\subsection{Velocity-gauge real-time TDDFT}
The velocity-gauge (VG) formulation of real-time TDDFT (VG-RT-TDDFT) which is well suited for the treatment of laser-driven electron dynamics in Bloch-periodic systems has been previously discussed by several authors~\cite{Yabana1996,Yabana2012,Krieger2015,Pemmaraju2017}. In particular, the numerical  approach employed in this work, which is based on a linear-combination of atomic-orbitals (LCAO) implementation of the generalized Kohn-Sham (GKS)~\cite{Baer2018} real-time TDDFT  framework has been described in detail elsewhere~\cite{Pemmaraju2017,Pemmaraju2018b}. Within this approach the time-dependent generalized Kohn-Sham (TDGKS) equations for electron dynamics 
\begin{eqnarray}\label{vgtdks}
&\imath\hbar\frac{\partial}{\partial t}{\phi}_i(\overrightarrow{r},t)=\hat{{H}}_{GKS}{\phi}_i(\overrightarrow{r},t)
\end{eqnarray} 
with ${\phi}_i$ being single-particle orbitals and $\hat{{H}}_{GKS}$ the VG GKS Hamiltonian, are propagated in real-time. The VG GKS Hamiltonian is given by

\begin{eqnarray}\label{HGKS}
\hat{{H}}_{GKS}= &\frac{1}{2m}\left[  \overrightarrow{\hat{p}} +\frac{e}{c}\overrightarrow{A}(t)\right]^2  + \hat{\tilde{V}}_{ion}+ \nonumber \\ &\int d\overrightarrow{r}^\prime \frac{e^2}{|\overrightarrow{r}-\overrightarrow{r}^\prime|}n(\overrightarrow{r}^\prime,t) + \hat{V}_{XC}[\rho(\overrightarrow{r},\overrightarrow{r}',t)]
\end{eqnarray}

and features a kinetic term that includes coupling to the time-dependent vector potential $\overrightarrow{A}(t)$ representing the applied external fields, the VG electron-nuclear interaction term $\hat{\tilde{V}}_{ion}$, the Hartree potential and a non-multiplicative XC operator $\hat{V}_{XC}$ as a functional of the instantaneous single-particle GKS density matrix $\rho(\overrightarrow{r},\overrightarrow{r}',t)$~\cite{Baer2018}.  Time propagation of equation~\ref{vgtdks} yields at every time-step, relevant quantities such as the GKS 1-body density matrix
\begin{equation*}
\rho(\overrightarrow{r},\overrightarrow{r}',t)=\sum_{i}^{N_{occ}} {\phi}_i(\overrightarrow{r},t){\phi}_i^*(\overrightarrow{r}',t)
\end{equation*}
electron density,
\begin{equation*}
n(\overrightarrow{r},t)=\rho(\overrightarrow{r},\overrightarrow{r},t)=\sum_{i}^{N_{occ}}|\tilde{\psi}_i(\overrightarrow{r},t)|^2
\end{equation*}
and macroscopic current
\begin{equation}\label{macI}
\overrightarrow{I}(t)=-\frac{e}{\Omega}\int_{\Omega}d\overrightarrow{r}\overrightarrow{j}(\overrightarrow{r},t).
\end{equation}
which in turn is obtained from the time-dependent current density $\overrightarrow{j}(\overrightarrow{r},t)$ given as
\begin{equation}
\overrightarrow{j}(\overrightarrow{r},t)=\sum_{i}\frac{e}{2m} \left\lbrace {\phi}^*_i(\overrightarrow{r},t) \overrightarrow{\pi}{\phi}_i(\overrightarrow{r},t) + c.c \right\rbrace
\end{equation}
and includes the generalized momentum
\begin{equation}
\overrightarrow{\pi} = \frac{m}{\imath\hbar} [\overrightarrow{r}, \hat{{H}}_{GKS}].
\end{equation}
Following numerical time propagation of the TD GKS equation~\ref{vgtdks}, frequency domain  observables related to the time-dependent electron density or current are readily calculated via Fourier transformation\cite{Marques2012a,Bertsch2000,Yabana2012,Pemmaraju2017}.
\subsection{Tuned range-separated hybrid functionals}
As outlined in the introduction, one of the aims of this work is to investigate in the context of extended systems, the use of a range-separated hybrid (RSH) exchange-correlation (XC) functional form that provides improved accuracy compared to semi-local approximations with regards to the treatment of both core and valence electrons simultaneously. To this end an RSH functional which combines both long-range (LR) and short-range (SR) corrections is employed according to the following strategy: For an adequate treatment of valence excitations including excitonic effects, LR corrections along the lines proposed by Refaely-Abramson et al~\cite{Refaely-Abramson2015a} and previously demonstrated in the case of monolayer h-BN~\cite{Pemmaraju2018b} are employed.  Additional SR corrections which primarily affect core-electron energies are then applied based on satisfying an approximate Koopman condition~\cite{Dabo2010} where by core-level GKS eigenenergies are required to match core-electron removal energies from $\Delta$ self-consistent field ($\Delta$SCF)~\cite{Jones1989} calculations.  Such a procedure is analogous in the case of finite systems to enforcing  the linearity condition for orbital energies (LCOE)  as outlined previously by a number of authors~\cite{Imamura2015,Dabo2010,Nguyen2018}. 

To incorporate SR and LR corrections, the following partitioning of the Coulomb operator is employed featuring the error- and complementary-error functions:
\begin{eqnarray}\label{RSHC}
\fl
\frac{1}{|\overrightarrow{r}-\overrightarrow{r}^\prime|} = & \frac{ \alpha + \beta_{LR}~\mathrm{erf}(\omega_{LR} |\overrightarrow{r}-\overrightarrow{r}^\prime| ) + \beta_{SR}~\mathrm{erfc}(\omega_{SR} |\overrightarrow{r}-\overrightarrow{r}^\prime| )}{|\overrightarrow{r}-\overrightarrow{r}^\prime|} + \nonumber \\ & \frac{1-\lbrace \alpha + \beta_{LR}~\mathrm{erf}(\omega_{LR} |\overrightarrow{r}-\overrightarrow{r}^\prime| )\rbrace +\beta_{SR}~\mathrm{erfc}(\omega_{SR} |\overrightarrow{r}-\overrightarrow{r}^\prime| )}{|\overrightarrow{r}-\overrightarrow{r}^\prime|}
\end{eqnarray}
The first and second terms on the right hand side respectively are then used to evaluate the nonlocal Hartree-Fock (HFX) like and local density approximation (LDAX) like contributions to the exchange energy. This leads to an effective partitioning of the total XC energy as
\begin{eqnarray}\label{ERSH}
E_\mathrm{XC} = &(\alpha + \beta_{SR}) E^\mathrm{SR}_\mathrm{HFX} + (1-\alpha-\beta_{SR}) E^\mathrm{SR}_\mathrm{LDAX} + \nonumber \\ & (\alpha+\beta_{LR})E^\mathrm{LR}_\mathrm{HFX} + (1-\alpha-\beta_{LR})E^\mathrm{LR}_\mathrm{LDAX}
+E_\mathrm{LDAC}
\end{eqnarray}
where the $E^\mathrm{SR}_\mathrm{HFX}, E^\mathrm{LR}_\mathrm{HFX}$ respectively stand for the short-range (SR) and long-range (LR) Hartree-Fock exchange (HFX) contributions with similar notation for the LDA counterparts. The correlation contribution to the total energy $E_\mathrm{LDAC}$ is set to be that of the LDA~\cite{Perdew1981}.  Thus the functional form adopted features fractions $(\alpha + \beta_{SR}),(\alpha+\beta_{LR})$ of HFX in the SR and LR respectively. For brevity, the functional form implied by eqns.~\ref{RSHC},~\ref{ERSH} is henceforth refereed to as the short- and long-range corrected (SLRC) form. The SR and LR contributions to LDAX are calculated using the Coulomb attenuated form due to Toulouse \textit{et al}~\cite{Toulouse2004}. The parameters $\alpha, \beta_{LR}, \omega_{LR}$ are chosen such that 
\begin{eqnarray}\label{LRXC}
\alpha + \beta_{LR} = \frac{1}{\epsilon_{\infty}}
\end{eqnarray}
which ensures the correct long-range  $\frac{1}{\epsilon_{\infty}r}$ behavior of the XC potential in the $r\rightarrow\infty$ asymptotic limit while tuning $\alpha, \omega_{LR}$ so that the GKS band-gap of the extended system matches the quasiparticle band-gap obtained from benchmark GW calculations~\cite{Ferreira2019} 
\begin{figure}[htbp]
	\centering
	\captionsetup{width=\linewidth}
	\includegraphics[width=0.5\textwidth]{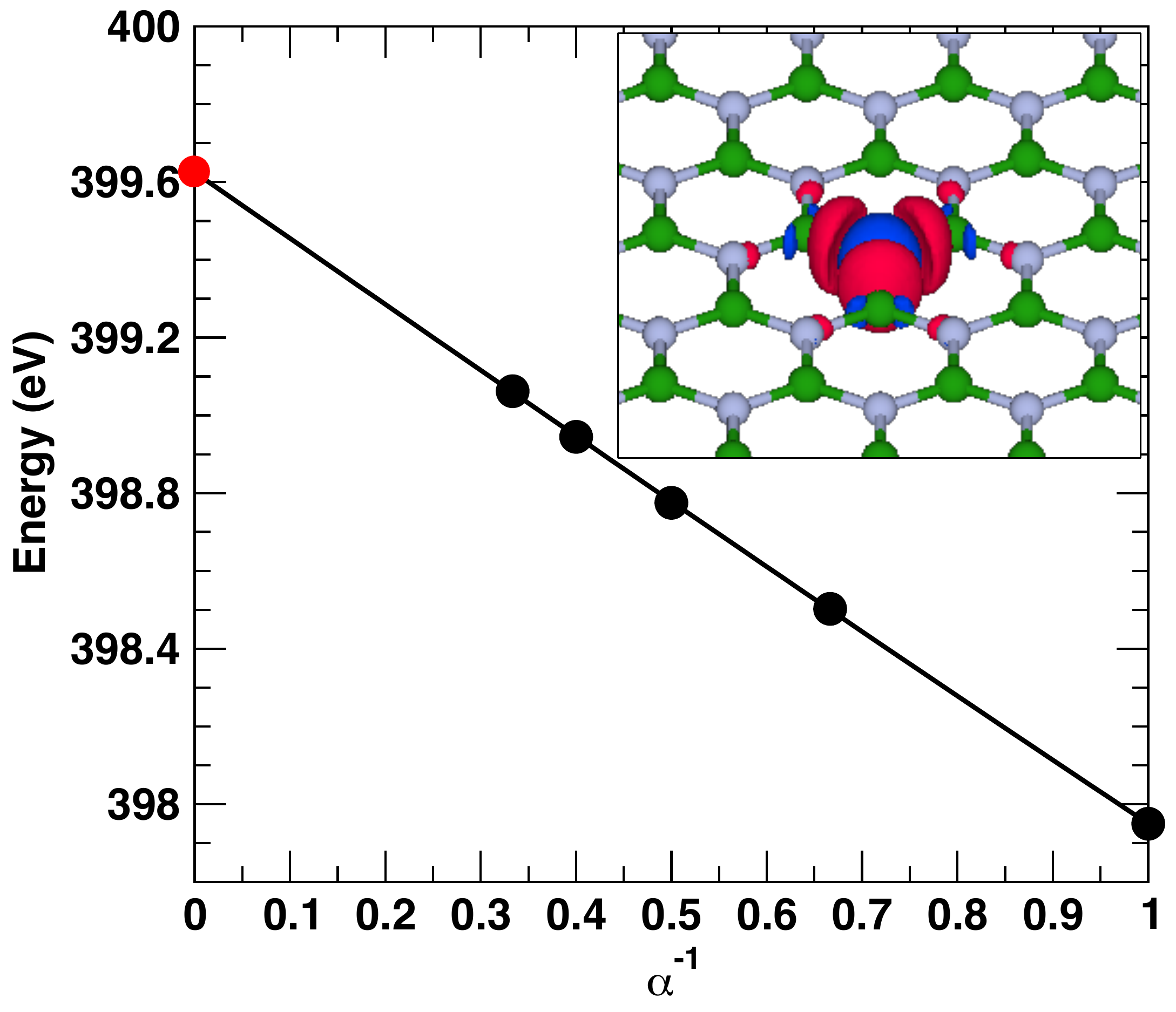}
	\caption{Charged supercell calculation of the absolute binding energy in eV (black line) of the N $1s$ core-level in 2D h-BN with extrapolation to the infinite supercell size limit under the scaling L $\rightarrow \alpha$ L of linear cell dimensions. $\alpha=1$ corresponds to a 18.927$\times$16.392$\times$18.927 a.u$^{3}$ orthorhombic supercell. $\alpha^{-1} \rightarrow 0$ corresponds to the relevant infinite cell size limit and the extrapolated binding energy is indicated by the red circle. (inset) Isosurface depicting the charge density difference between a N $1s$ core-ionized configuration and the neutral ground state of 2D h-BN. Green and blue spheres represent B and N sites. Structure visualization is performed using VESTA 3~\cite{Momma2011}} 
	\label{cip}  
\end{figure}
The choice of $\beta_{SR}, \omega_{SR}$ on the other hand predominantly determines the core-level GKS eigenenergies.  In principle both $\beta_{SR}$ and $\omega_{SR}$ can be tuned separately so that the following condition is satisfied for a chosen set of core-levels: 
\begin{eqnarray}\label{Ecore}
\varepsilon^{A}_{c} = -( E^{A+}_{c} - E^{GS})
\end{eqnarray} 
where $\varepsilon^{A}_{c}$ is the GKS eigenvalue of the core-level $c$  on atom $A$,  $E^{GS}$ is the total energy of the neutral ground state of the system of interest and $E^{A+}_{c}$ is the total energy of the positively ionized system resulting from the removal of one electron form the core-orbital $c$ localized on atom A. Thus $( E^{A+}_{c} - E^{GS})$ represents a core-electron removal energy including final-state electronic screening effects which can be accessed from a constrained-DFT $\Delta$SCF~\cite{Gilbert2008} or a quasiparticle GW calculation~\cite{Golze2018}.  In this work, $( E^{A+}_{c} - E^{GS})$ is calculated as a LDA total-energy difference for 2D periodic h-BN using a supercell  $\Delta$SCF approach including 2D image charge corrections to $E^{A+}_{c}$ from finite size scaling~\cite{Cerioni2012, Komsa2013}(see Fig.~\ref{cip}). Because of the localized nature of the core hole and associated short-range screening response (see Fig. ~\ref{cip})  supercell total energy methods for charged defects can be  used to approximate $E^{A+}_{c}$~\cite{Komsa2013,Ozaki2017}.  Since in h-BN, the B $1s$ and N $1s$ orbitals represent the core-states of interest, the removal energy $( E^{A+}_{c} - E^{GS})$ is calculated separately for each core hole case .i.e., $c=1s$ for $A=$B, N (see Fig.~\ref{cip} and Table ~\ref{fits}). 

\begin{table}[htbp]
	\begin{center}
		\begin{adjustbox}{width=0.75\textwidth}
			\begin{tabular}{ |c|c|c|c|}
				\hline
				property & LDA eigenvalue (eV) &SLRC eigenvalue (eV) &  Target energy (eV) \\
				\hline
				-$\varepsilon^{B}_{1s}$ & 174.74 &190.38  & 192.18 \\
				\hline
				-$\varepsilon^{N}_{1s}$ & 375.54 &404.85& 399.62\\
				\hline
				$\Delta_{G}^{\Gamma}$ & 6.08 &8.94 & 9.07~\cite{Ferreira2019}\\
				\hline
			\end{tabular}
		\end{adjustbox}
	\end{center}
	\caption{ Core-electron binding energies and the band-gap at $\Gamma$ ($\Delta_{G}^{\Gamma}$) inferred from KS LDA or GKS SLRC functional eigenvalues are compared against target energies inferred from  $\Delta$SCF or GW~\cite{Ferreira2019} calculations. Within the GKS scheme~\cite{Seidl1996}, the SLRC functional eigenvalues reproduce the target single-particle excitation energies for both core- and valence-electrons to within 2\% for the tuned parameter set $\alpha=0$, $\beta_{LR}=1$, $\omega_{LR}=0.129~\mathrm{Bohr^{-1}}$, $\beta_{LR}=1$, $\omega_{SR}=1.40~\mathrm{Bohr^{-1}}$. The standard KS system is not required to reproduce single-particle excitation energies except for the highest occupied orbital~\cite{Perdew2017}. }\label{fits}
\end{table}

The parameters on the right hand side of eqn.~\ref{RSHC}  are then determined as follows: In monolayer h-BN, the long-range asymptotic limit of the macroscopic dielectric constant $\epsilon_{\infty}^{||} \rightarrow 1$~\cite{Huser2013} dictates a choice of  $\alpha + \beta_{LR} = 1$ from eqn.~\ref{LRXC} above.  Physical considerations similarly suggest unscreened Fock exchange in the very short-range~\cite{Chen2018} which requires  $\alpha + \beta_{SR}=1$. While $\alpha$ is in principle a free parameter, in this work we set $\alpha=0$ to be consistent with a previously reported~\cite{Pemmaraju2018b} valence-electron LRC functional parameterization for h-BN without explicit treatment of core-electrons. Thus the effect of including core-electrons in the description is intended to be captured by $\beta_{SR}$ and the range-separation parameters $\omega_{SR}, \omega_{LR}$. The choice of $\alpha=0$ fixes both $\beta_{LR} =1$, $\beta_{SR}=1$ leaving $\omega_{SR},\omega_{LR}$ as the two parameters to be tuned in eqn.~\ref{RSHC}.  The two core-electron energy eigenvalues $\varepsilon^{B}_{1s}$, $\varepsilon^{N}_{1s}$ and the $\Gamma \rightarrow \Gamma$ GKS band-gap $\Delta_{G}^{\Gamma}$  in 2D h-BN represent three quantities to be fit to the corresponding quasiparticle energy differences as outlined above (see eqn.~\ref{Ecore}) by minimizing the percentage error in a least squares sense.  The resulting values for $\varepsilon^{B}_{1s}$, $\varepsilon^{N}_{1s}$, $\Delta_{G}^{\Gamma}$ and their comparison to target data are shown in Table~\ref{fits} for the optimized parameter set $\alpha=0$, $\beta_{LR}=1$, $\omega_{LR}=0.129~\mathrm{Bohr^{-1}}$, $\beta_{LR}=1$, $\omega_{SR}=1.40~\mathrm{Bohr^{-1}}$. 

\begin{figure}[htbp]
	\centering
	\includegraphics[width=\textwidth]{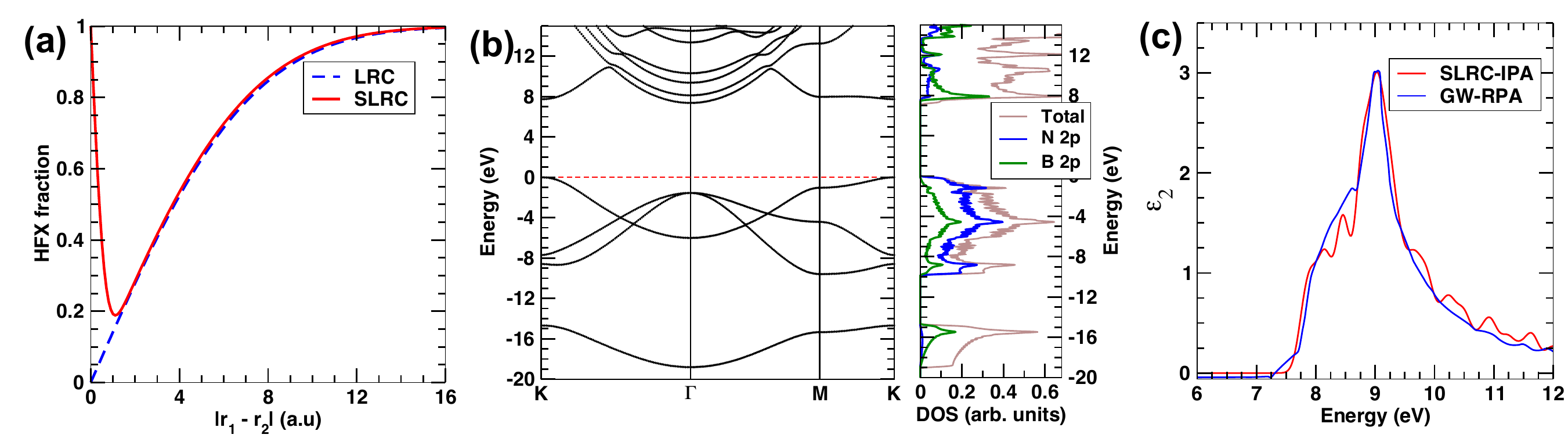}\caption{(a) Variation of the fraction of Hartree-Fock exchange (HFX) as a function of inter-electron separation $|\mathbf{r}_1-\mathbf{r}_2|$ in the SLRC functional form used in this work is plotted (red line). The HFX fraction in a previously reported~\cite{Pemmaraju2018b} valence-only tuned LRC  functional for 2D h-BN when core-electrons are not included explicitly in the description is shown (blue dashed line). Beyond 1 a.u, the SLRC and LRC functionals are very similar while non-local exchange effects pertaining to core-electrons are  captured in the SLRC functional through the inclusion of HFX in the very-short range. (b) Electronic band structure and density of states (DOS) of 2D h-BN calculated using the SLRC functional. (c) Optical response of 2D h-BN within the independent particle approximation (IPA) calculated using the SLRC approach in an LCAO basis is compared with GW RPA response from the literature~\cite{Ferreira2019}.}  
	\label{ipa}
\end{figure}

The variation in the effective fraction of Fock exchange as a function of inter-electron separation for the tuned SLRC functional is plotted in Figure~\ref{ipa}. The corresponding GKS band structure and independent-particle approximation (IPA) optical response of 2D h-BN are also shown. Thus the SLRC functional yields a description in which both core and valence  electron GKS eigenvalues approximate single-particle addition/removal energies of the system.  In the next section RT-TDDFT results for both equilibrium and non-equilibrium optical response employing this SLRC functional are discussed.

\subsection{Numerical details}
Numerical simulations in this work are carried out using a linear-combination of atomic-orbital (LCAO) framework for Velocity-gauge real-time TDDFT (VG-RT-TDDFT)~\cite{Pemmaraju2017,Pemmaraju2018b} based on a modification of the SIESTA~\cite{Soler2002} density functional platform. Accordingly, for 2D h-BN, electron-ion interactions are modeled through norm-conserving nonrelativistic LDA pseudopotentials generated using the Troullier-Martins~\cite{Troullier1991} scheme while the GKS Hamiltonian, density matrix and wavefunctions are represented in an LCAO basis. Non-local Fock exchange integrals are evaluated using an auxiliary Gaussian basis set representation as described previously~\cite{Pemmaraju2018b}. Pseudopotential and basis set parameters are listed in Table~\ref{hbncomp}. Brillouin zone sampling is carried out using $\Gamma$-centered $\mathbf{k}$-point grids also as listed in Table~\ref{hbncomp}. Time propagation of the GKS~\cite{Baer2018,Pemmaraju2018b} equations is carried out using a Crank-Nicholson scheme~\cite{Crank1947} with a predictor-corrector step.
\begin{table}[htbp]
	\begin{center}
		\begin{adjustbox}{width=\textwidth}
			\begin{tabular}{ |c|c|c| }
				\hline
				\multicolumn{3}{|c|}{LCAO computational parameters for 2D h-BN simulations} \\
				\hline
				lattice constant (\AA)  &  \multicolumn{2}{|c|}{2.504} \\
				\hline
				Pseudopotential valence configuration & B: 1$s^2$,2$s^2$,2$p^1$&N:1$s^2$,2$s^2$,2$p^3$\\
				\hline
				\multirow{2}{*}{basis set ($nl$-$\zeta$)} & B: $1s$-2,$2s$-2,$2p$-2,$3d$-1  & N:$1s$-2,$2s$-2,$2p$-2,$3d$-1   \\
				&  8$\times ghost:2s$-1&    - \\
				\hline
				Real-space mesh-cutoff & \multicolumn{2}{|c|}{940 Ry} \\
				\hline
				DFT SCF $\mathbf{k}$-point grid &\multicolumn{2}{|c|}{$\Gamma-24\times24\times1$}\\
				\hline
				VG-RT-TDDFT $\mathbf{k}$-point grid & \multicolumn{2}{|c|}{$\Gamma-24\times24\times1$}\\
				\hline
				VG-RT-TDDFT time step  & \multicolumn{2}{|c|}{0.0075 a.u} \\
				\hline
			\end{tabular}
		\end{adjustbox}
	\end{center}
	\caption{Computational parameters used for valence and core excitation simulations of 2D h-BN. The LCAO basis set is specified using $nl$-$\zeta$ notation where $n,l$ are principal and azimuthal quantum numbers respectively and $\zeta$ represents the number of functions of each $nl$ type. Additional $ghost$ basis functions of $s$ character are used in the vacuum region surrounding the 2D h-BN layer within the LCAO approach. Hartree and semilocal exchange-correlation potentials are evaluated by representing the charge density over a real-space mesh~\cite{Soler2002} for which an equivalent planewave  cutoff is listed.}\label{hbncomp}
\end{table}

\section{Results}
\begin{figure}[htbp]
	\centering
	\includegraphics[width=\textwidth]{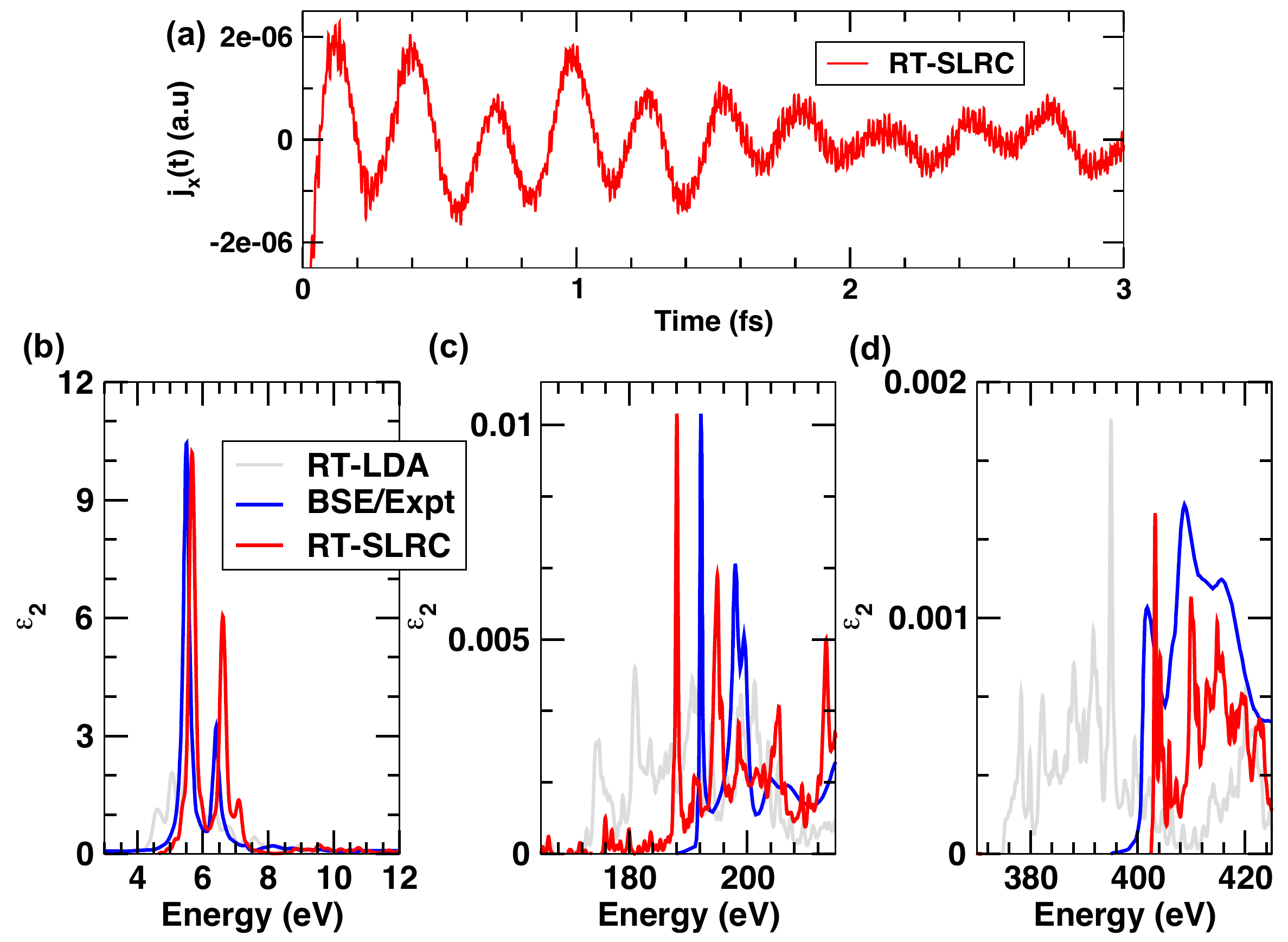}\caption{(a) Time evolution of the in-plane macroscopic current in 2D h-BN following a weak delta electric field perturbation calculated using the SLRC hybrid functional. The current fluctuations include low and high frequency components that correspond to valence and core-level dipole excitations respectively.  (b) Valence optical absorption spectrum in 2D h-BN from the RT-SLRC approach (red) compared to BSE spectra (blue) digitized from reference~\cite{Ferreira2019}.  (c) Boron and (d) Nitrogen K-edge X-ray absorption spectra in 2D h-BN from the RT-SLRC approach (red) compared with reference experimental data (blue) digitized from references~\cite{Wang2015} and ~\cite{McDougall2017} respectively.  Adiabatic LDA (RT-LDA) spectra  are also shown (gray) for both the valence and core-excitation cases.}  
	\label{linr}
\end{figure}
\begin{table}[htbp]
	\begin{center}
		\begin{adjustbox}{width=0.75\textwidth}
			\begin{tabular}{ |c|c|c|c|}
				\hline
				 transition & RT-SLRC  (eV) &Reference (eV) &  $\%$ error \\
				\hline
                 valence exciton  & 5.67 & 5.58~\cite{Ferreira2019} & 1.6 \\
				\hline
				 B $1s \rightarrow \pi^*$  & 188.1 & 192.1~\cite{Wang2015}  & 2.0 \\
				\hline
				 N $1s \rightarrow \pi^*$  & 403.3 & 401.2~\cite{Wang2015}-401.8~\cite{McDougall2017} & 0.4\\
				\hline
			\end{tabular}
		\end{adjustbox}
	\end{center}
	\caption{ Energies of key valence and core-excitations in 2D h-BN involving as obtained from RT-SLRC TDDFT are compared with reference values. The $\%$ error in absolute RT-SLRC energies is the range of 2$\%$.}\label{lrtab}
\end{table}
\subsection{Linear response}
To evaluate the tuned SLRC functional, firstly the response of 2D h-BN to a weak Dirac-$\delta$ like time-localized electric field perturbation is simulated and fluctuations of the induced macroscopic current are calculated in real-time as plotted in Figure~\ref{linr}(a).  The corresponding frequency domain dielectric response is obtained through Fourier transformation of the time-dependent macroscopic current~\cite{Yabana2012}.  The low and high frequency fluctuations apparent in the time-evolution of the current respectively correspond to the valence- and core-electronic optical response. The imaginary part $\epsilon_2$ of the dielectric function is plotted in figure~\ref{linr} over energy ranges relevant to valence as well as B, N $1s$ (K-edge) core-excitations in h-BN.  For the sake of comparison, real-time adiabatic LDA (ALDA) results are also shown. As has been extensively reviewed previously~\cite{Byun2020,Onida2002}, Kohn-Sham ALDA does not capture excitonic features in the optical response either in the valence or core-excitation case~\cite{Pemmaraju2018b}. In  ALDA the energy onset of UV absorption in 2D h-BN is underestimated by over 1 eV and the strong excitonic enhancement of absorption at 5.58 eV predicted by benchmark BSE simulations~\cite{Ferreira2019} is not observed. In the absence of explicit self-interaction corrections~\cite{Perdew1981}, the ALDA XC potential asymptotically decays exponentially which translates to an over-screening of the Coulomb interactions necessary to stabilize excitons.  In contrast because of the incorporation of the correct $\frac{1}{\epsilon_{\infty}r}$ long-range asymptotic behavior of the XC potential via the screened exchange component of the nonlocal SLRC functional, the RT-SLRC time dynamics reproduces the strong valence excitonic features expected in 2D h-BN to within 0.2 eV of BSE results. In the case of X-ray excitations at the B and N K-edges in the hundreds of eV range, ALDA once again underestimates the onset of the absorption edge by over $9\%$ and $6\%$ respectively while also missing  prominent core-excitonic features leading to an overall spectral line shape that bears little resemblance to experimental spectra. The short-range component of the nonlocal Fock-like exchange incorporated into the SLRC functional overcomes these shortcomings. Firstly, it improves absolute core-excitation energies by mitigating delocalization errors on the length-scale of localized core electrons towards restoring the piece-wise linearity condition on core-orbital energies~\cite{Imamura2015}. Concomitantly,  because of  reduced screening of short-range electron-hole interactions relative to ALDA, the density evolution under the SLRC XC potential encodes core-excitonic effects. As apparent from Fig.~\ref{linr}(b), \ref{linr}(c) and table~\ref{lrtab}, with the RT-SLRC approach absolute energies of the B, N K-edge positions are reproduced to within 2$\%$ of experiment. In particular the excitonic $1s \rightarrow \pi^{*}$ transitions at the B and N K-edges are correctly described while overall spectral features in a 20 eV energy range above the onset are also in reasonable agreement with experiment considering that the frequency-dependence of lifetime broadening effects is not included within the adiabatic XC approximation used here. Thus with time-propagation of the GKS equations for a delta perturbation, the tuned SLRC XC potential employed in this study is able to provide a balanced description of  electron-hole excitations across a wide energy range in strongly excitonic 2D h-BN.  

\subsection{Attosecond Transient X-ray absorption} Attosecond Transient Absorption Spectroscopy (ATAS) ~\cite{Geneaux2019,Kraus2018a,Schultze2014,Picon2019,DeGiovannini2013,Sato2018} typically involves a pump-probe protocol wherein the absorption of an ultrashort probe pulse in a material excited by a pump pulse is measured. The intensity, duration and shape of the pump pulse as well as the pump-probe delay provide external control parameters that can be tuned. The simulation,  within a TDDFT framework, of ATAS employing low energy IR-XUV probe pulses in molecules~\cite{DeGiovannini2013,Nguyen2016} and solids~\cite{Lucchini2016a,Sato2018}  has been discussed in the literature. Soft X-ray transient absorption (XTAS)~\cite{Geneaux2019,Kraus2018a, Picon2019} is similar in principle except that higher frequency (100 - 1000 eV) wide band-width X-ray probe pulses tuned to core-excitation edges are typically employed yielding an element specific perspective of transient electron dynamics with attosecond time-resolution  Within the real-time velocity gauge formulation of TDDFT~\cite{Yabana2012}, one has access to the time-dependent macroscopic current $\overrightarrow{I}(t)$ (Eqn.~\ref{macI}) within a material that is under the influence of external laser fields described by the vector potential 
\begin{equation}\label{Vecp}
\overrightarrow{A}(t)=-c \int^{t} \overrightarrow{E}(t') dt' 
\end{equation}
obtained from an electric field $\overrightarrow{E}(t)$. In the frequency domain, the macroscopic current $\overrightarrow{I}(\omega)$ associated with a weak probe pulse considered to be in the linear response regime and described by a given external perturbation $\overrightarrow{E}(\omega)$ can be written as 
\begin{equation}\label{iw}
\overrightarrow{I}(\omega) = \overline{\overline{\sigma}}(\omega) . \overrightarrow{E}(\omega)
\end{equation}
where $\overline{\overline{\sigma}}$ is the conductivity tensor.  

In a pump-probe scheme for a certain choice of pump and probe fields described by a field profile $\overrightarrow{E}_{pump\mh probe}(t)$ one can calculate through propagating the GKS equation~\ref{vgtdks} the associated macroscopic current $\overrightarrow{I}_{pump\mh probe}(t)$ using equation.~\ref{macI}. Similarly, for the pump field $\overrightarrow{E}_{pump}(t)$ alone, once can calculate the corresponding current $\overrightarrow{I}_{pump}(t)$. Then the probe current in the pump-probe scheme is defined as
\begin{equation}\label{ipr}
\overrightarrow{I}_{probe}(t) \equiv \overrightarrow{I}_{pump\mh probe}(t) - \overrightarrow{I}_{pump}(t)
\end{equation}
This separation of the total current into pump and probe parts is sensible for  probes of any intensity but $\overrightarrow{I}_{probe}(t)$ has a dependence on the parameters describing the pump pulse such as its shape, intensity, duration, polarization etc indicated collectively by the set $\mathcal{P}$ as well as the pump-probe delay $\Delta T$.  Additionally, for weak probe pulses, equation~\ref{iw} can be generalized to
\begin{equation}\label{iprw}
\overrightarrow{I}_{probe}(\omega) = \overline{\overline{\sigma}}_{\mathcal{P}}(\omega, \Delta T) . \overrightarrow{E}_{probe}(\omega)
\end{equation} 
in terms of the \textit{transient} frequency dependent conductivity tensor $\overline{\overline{\sigma}}_{\mathcal{P}}(\omega, \Delta T)$ which now depends on the pump parameters $\mathcal{P}$ and the delay $\Delta T$. For comparison to XTAS experiments, it is useful to calculate the \textit{transient} dielectric tensor
\begin{equation}\label{epsw}
 \overline{\overline{\epsilon}}_{\mathcal{P}}(\omega, \Delta T)  =  1 + \frac{4\pi\imath\overline{\overline{\sigma}}_{\mathcal{P}}(\omega, \Delta T)}{\omega}
\end{equation} 
whose imaginary part is related to X-ray absorption. 
\begin{figure}[htbp]
	\centering
	\includegraphics[width=0.75\textwidth]{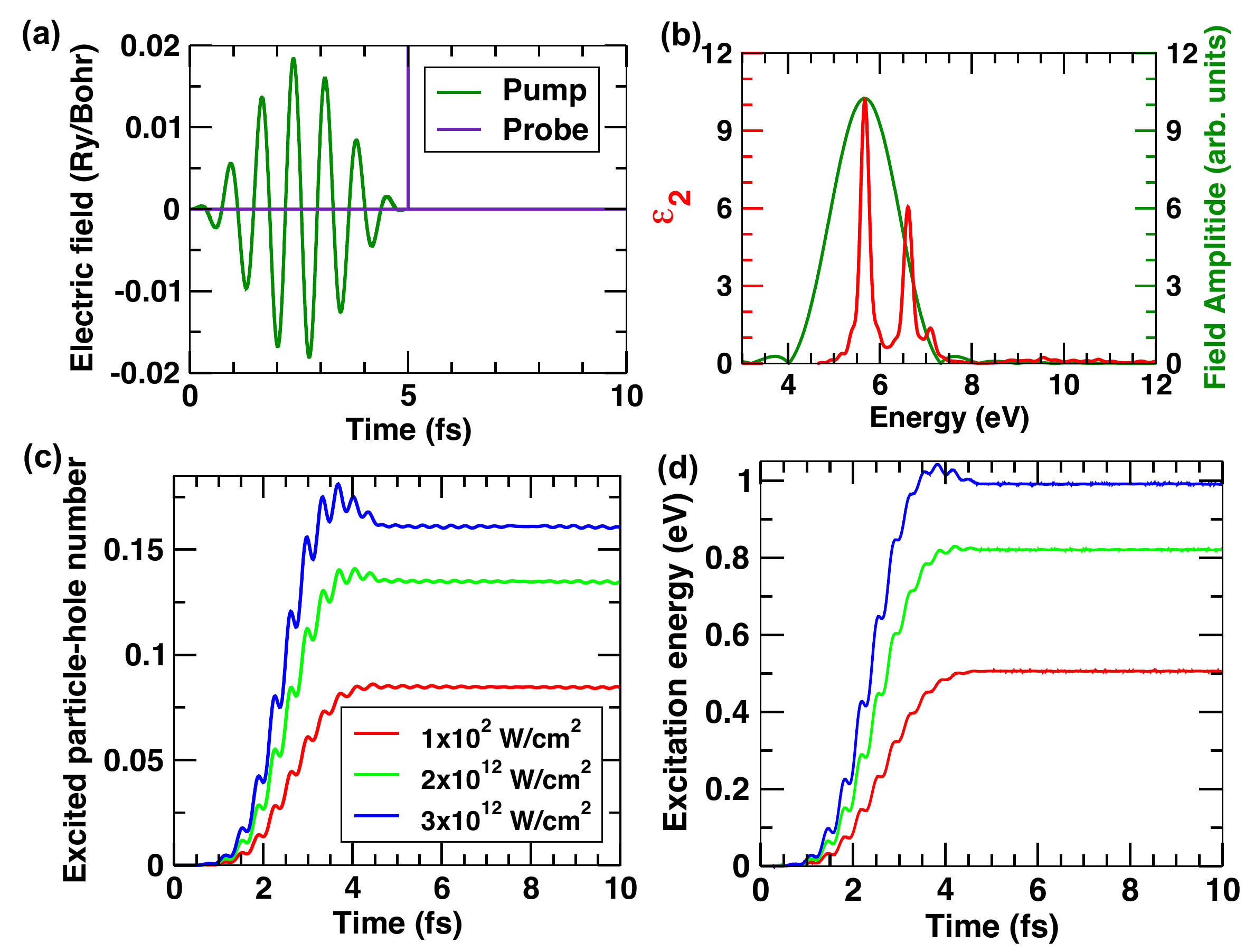}\caption{Pump-probe XTAS setup simulated in this work. (a) A 5 fs pump laser pulse with a $Sin^2$ envelope function and in plane polarization is followed by a delta perturbation probe (purple) polarized normal to the pump but including in-plane and out-of-plane components. (b) Frequency profile of the pump pulse (green) centered at 5.67 eV with a $\sim$1.7 eV FWHM is shown along with the UV optical absorption spectrum of 2D h-BN (red). (c) Number of particle-hole pairs per unitcell created as a function of the peak intensity of the pump pulse.  (d) Energy per unit cell deposited in the material over the course of the interaction of the pump pulse with 2D h-BN shown for different peak intensities of the laser pulse. }  
	\label{ppset}
\end{figure}
To illustrate the simulation of XTAS using the RT-SLRC hybrid functional approach within GKS-TDDFT, the pump-probe set up shown in Fig.~\ref{ppset} is considered in this study. Accordingly a 5 femtosecond (fs) pump pulse with a  a $Sin^2$ envelope function and a central frequency of 5.67 eV tuned to the first valence exciton in 2D h-BN is employed. Because of the finite time duration of the pulse, its frequency band-width is roughly 2 eV. Different intensities of this pulse impinging on 2D h-BN with its electric field polarized in-plane are simulated. The chosen intensity range of $\sim10^{12}$ W/cm$^2$ corresponds to that typically employed~\cite{Schultze2014} in attosecond pump-probe experiments of electron dynamics. The pump pulse acts on the sample to excite electron-hole pairs at a density which is roughly 2\% of the valence electron density in 2D h-BN. Concomitantly, it also deposits on the order of 1 eV per unitcell of energy during the course of its interaction with the material. Immediately following the passage of the pump pulse, the material is probed using a weak delta perturbation. While a weak probe of any shape could be employed, a delta perturbation corresponds to the infinite band-width limit which is convenient in this context as it provides a probe-shape independent view of the transient dielectric function. The electric field of the probe is polarized along the diagonal of the plane perpendicular to the in-plane pump electric field. The probe therefore couples to both $1s \rightarrow \pi^{*}$ and $1s \rightarrow \sigma^{*}$ excitations. With this set up, the transient dielectric functions for different pump intensities are calculated and its imaginary part ($\mathrm{\varepsilon_2}$) is plotted in Fig.~\ref{ppxas} for both the B and N K-edges. 
\begin{figure}[htbp]
	\centering
	\includegraphics[width=\textwidth]{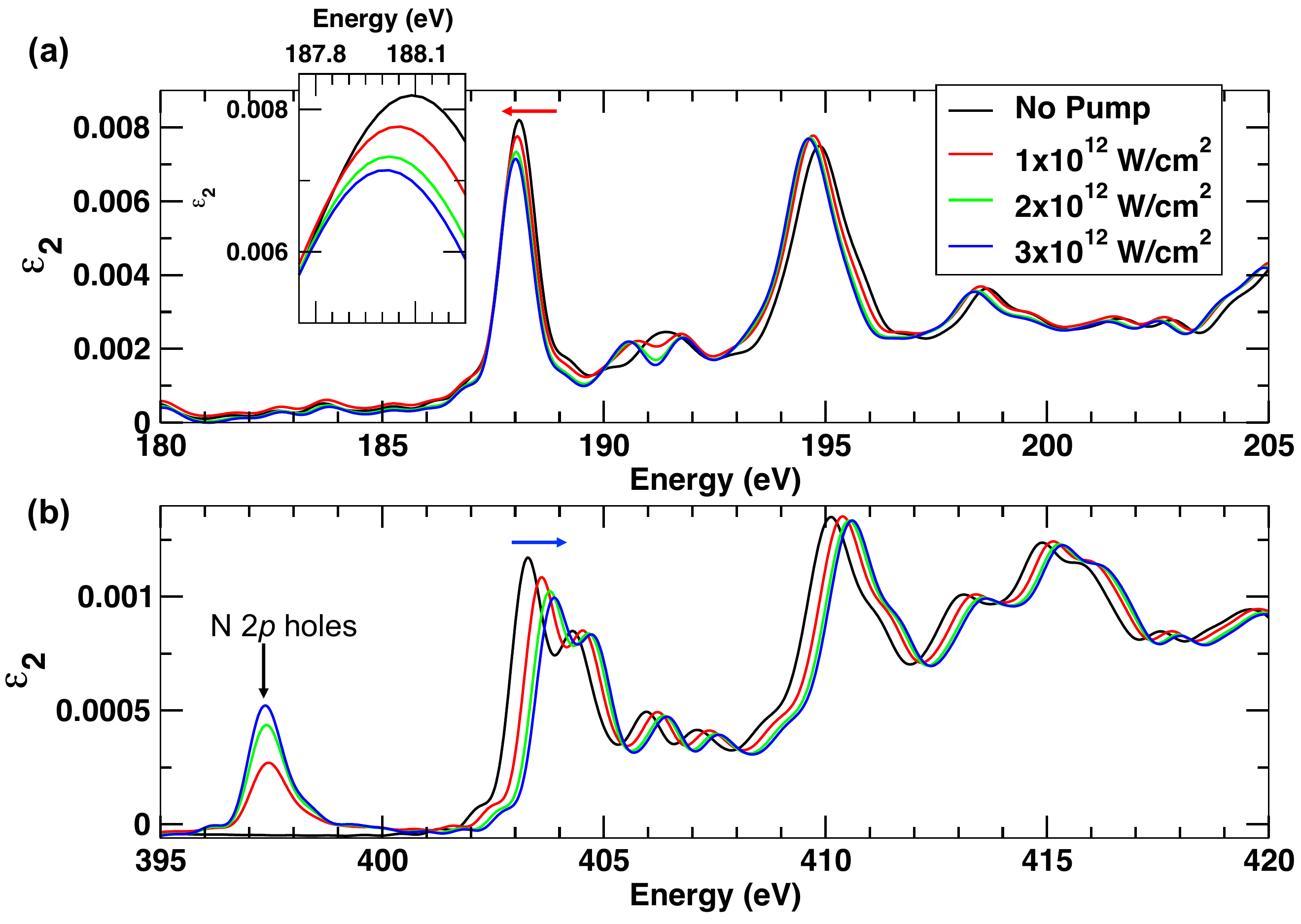}\caption{ RT-TDDFT X-ray Transient Absorption Spectra (XTAS) of 2D h-BN calculated using the SLRC hybrid functional and the pump-probe setup described in Fig~\ref{ppset}. (a) B K-edge XTAS for different intensities of the pump. (inset) Zoomed-in view of the spectra around 188.1 eV showing a transient red-shift of the $1s \rightarrow \pi^{*}$ peak(b) N K-edge XTAS for different intensities of the pump.}  
	\label{ppxas}
\end{figure}

\begin{figure}[htbp]
	\centering
	\includegraphics[width=0.75\textwidth]{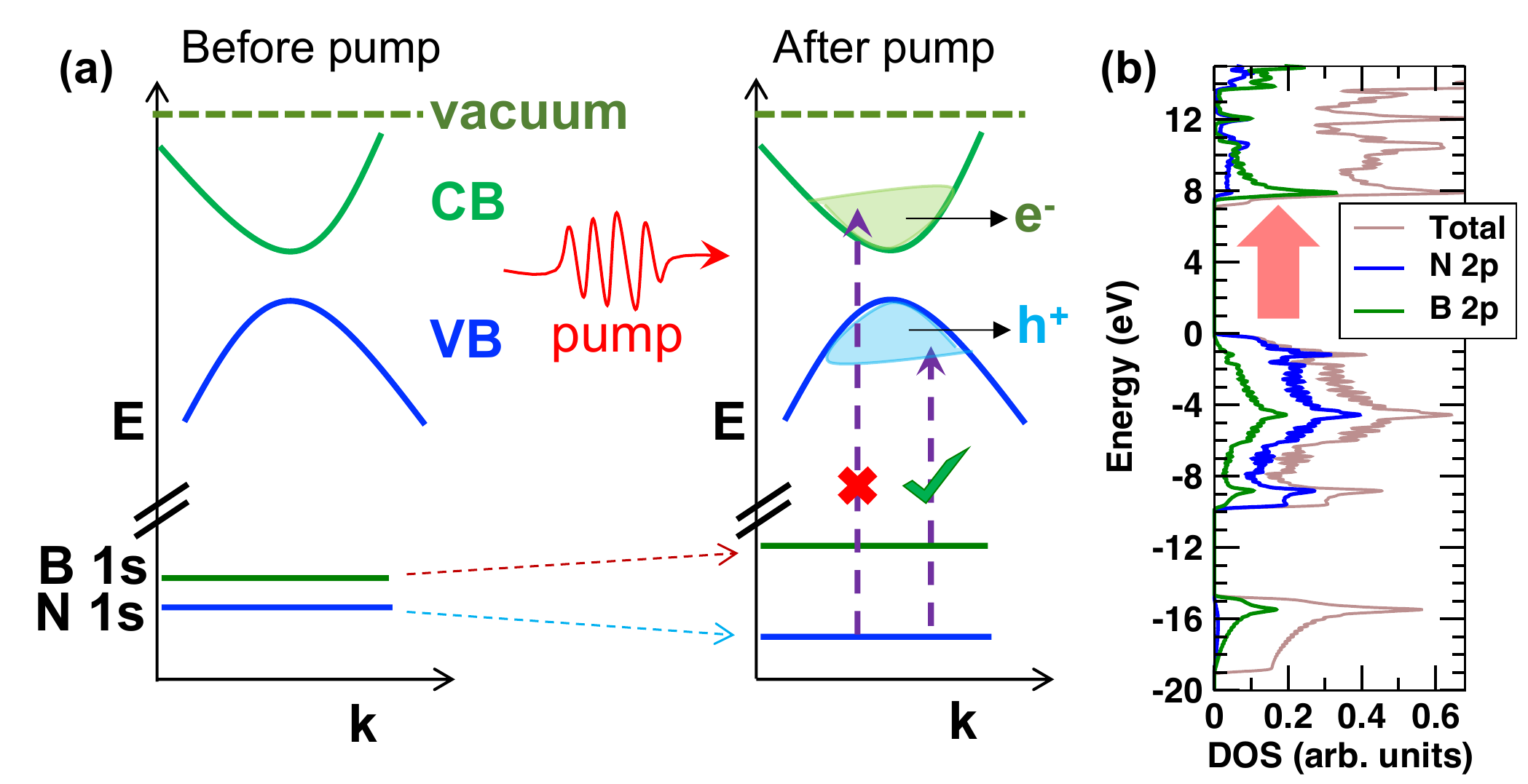}\caption{Schematic showing qualitatively the mechanisms of excitation induced state blocking and transient core-level shifts. (a) Valence band (VB), conduction band (CB) and core level band structure before and after excitation by a pump pulse. The pump pulse effectively creates an electron (hole) population in orbitals with conduction (valence) band character whether or not the electron-hole pairs form excitonic bound states. Then some core-excitations to  CB states can get transiently blocked while core-excitations to  VB states can become allowed. Core-level binding energies can also exhibit transient shifts based on the orbital characters of CB and VB states.  (b) Electronic  density of states of 2D h-BN showing that the VB is predominantly N-2$p$  while the CB is mainly B-2$p$ in character.}  
	\label{pauli}
\end{figure}

By comparing the \textit{transient} dielectric function (tr-$\mathrm{\varepsilon_2}$) after the pump with its ground state counterpart (gs-$\mathrm{\varepsilon_2}$) calculated with no pump, it is apparent that tr-$\mathrm{\varepsilon_2}$ is modulated in several ways at the B and N K-edges. Firstly, it is seen that relative to gs-$\mathrm{\varepsilon_2}$ the $1s \rightarrow \pi^{*}$ peak in tr-$\mathrm{\varepsilon_2}$ at both the B (188.1 eV) and N (403.3 eV) K-edge exhibits a reduction in intensity or bleaching that increases with increasing pump fluence. This effect can be understood as an excitation induced Pauli-blocking~\cite{Kamat1989,Manser2014,Schultze2014,Geneaux2019}. In 2D h-BN the valence band maximum (VBM) and conduction band minimum (CBM) are  primarily of N-2$p$ orbital derived $\pi$ and B-2$p$ orbital derived $\pi^{*}$ characters respectively (see Fig~\ref{pauli}). Valence excitations in the energy range of the pump pulse therefore exhibit $\pi \rightarrow \pi^{*}$ character with the electronic part of the electron-hole wavefunction being predominantly $\pi^{*}$ like~\cite{Ferreira2019}. This is equivalent within a projected single-particle picture to a transient filling of the $\pi^{*}$  CBM states~\cite{Yabana2012} following laser excitation by the pump. As the probe pulse that immediately follows the pump drives $1s \rightarrow \pi^{*}$ transitions, a certain fraction of these are blocked in proportion to the transient occupation of the $\pi^{*}$ CBM states. Thus the intensity of the $1s \rightarrow \pi^{*}$ peak in tr-$\mathrm{\varepsilon_2}$ is reduced. Note that the higher lying $1s \rightarrow \sigma^{*}$ peaks occurring near $\sim$194.8 eV and 410.1 eV at the B and N K-edges respectively, do not show such a reduction in intensity as the pump excitation does not lead to significant occupation of the conduction band $\sigma^{*}$ states. Meanwhile, the pump also transiently induces a hole population near the VBM which is of $\pi$ character. Hence from the perspective of the probe pulse certain $1s \rightarrow \pi$ transitions of core electrons to VBM hole states become allowed. These would otherwise be blocked in the ground state when the valence band is fully filled.  This kind of Pauli-unblocking of core-excitations is readily apparent as a new $\sim$ 397.5 eV peak observed in the N K-edge tr-$\mathrm{\varepsilon_2}$ spectrum which is missing in gs-$\mathrm{\varepsilon_2}$.  A similar feature is not observed at the B K-edge however because the B $1s \rightarrow \pi$ oscillator strength is much weaker relative to the B $1s \rightarrow \pi^{*}$ strength given that the VBM has very little B-2$p$ character. This difference between the B and N K-edges shows the orbital selectivity of features in tr-$\mathrm{\varepsilon_2}$. 

In addition to the intensity modulations discussed in the previous paragraph, tr-$\mathrm{\varepsilon_2}$ peaks also exhibit certain energy shifts on the order of 0.1-0.4 eV. In particular we note that the  B $1s \rightarrow \pi^{*}$ peak at 188.1 eV is  transiently red-shifted (see Fig~\ref{ppxas}) while the N $1s \rightarrow \pi^{*}$ peak at 403.3 eV is blue-shifted. Higher lying $1s \rightarrow \sigma^{*}$ peaks show similar trends being transiently red- and blue-shifted at the B and N K-edges respectively. These can be understood in terms of valence excitation induced core-level energy shifts. Note that in h-BN, because of the predominant N-2$p$ and B-2$p$ orbital characters of the VBM and CBM respectively, an across the band-gap excitation has a charge-transfer like character whereby N sites are incrementally oxidized and B sites are reduced. The amount of fractional oxidation or reduction scales with the density of electron-hole pairs created by the pump. Localized core-electronic levels on atoms are sensitive to the local oxidation state because of its effect on electronic screening of the attractive nuclear potential.  Thus core-electron energy-levels on transiently oxidized species (N) see their binding energies increase while those on transiently reduced species (B) find themselves at lower binding energies relative to the vacuum level (see Fig.~\ref{pauli}). This behavior is exploited in time-resolved X-ray photoemission spectroscopy~\cite{Siefermann2014} that directly tracks transient changes in core-electron binding energies. In XTAS simulated with RT-TDDFT we do not directly calculate the transient core-electron single particle energy shifts but nevertheless the signature of these binding energy shifts is observed in the probe induced core to valence dipole transition energies. Accordingly, pump intensity dependent peak shifts that modulate tr-$\mathrm{\varepsilon_2}$ are seen in Fig~\ref{ppxas}. 

 While the XTAS spectra in Fig~\ref{ppxas} are analyzed qualitatively in terms of pump excitation induced state blocking and core-level shifts, in general depending upon the excitation regime, transient particle-hole response in semiconductors and insulators incorporates additional effects such as band gap renormalization~\cite{Spataru2004, Roth2019} and nonequilibrium screening induced exciton energy shifts~\cite{Schultze2014, Zhao2020}. For the specific mode of excitation and the low 1-2\% particle-hole density considered here, these effects, to the extent that they are modeled within adiabatic GKS theory, do not seem prominent. The decomposition of the full electron-hole transient response available from XTAS into separate one-particle and two-particle contributions~\cite{Schultze2014} can in principle be carried out within a GKS RT-TDDFT framework by also simulating separately, time-resolved core and valence~\cite{DeGiovannini2017} photoemission spectra. Such simulations could enable a comparison of GKS RT-TDDFT to nonequilibrium quasi-particle theories~\cite{Perfetto2015a,Roth2019} and will be a subject of future studies. Nevertheless, the lack of explicit memory dependence in the XC potentials studied here also imposes certain limitations with regards to the description of excited state energy relaxation and decoherence~\cite{Wijewardane2005, Ullrich2012,Marques2012a}. Additionally, it should be noted that for time-overlapped pump and probe pulses transient spectra exhibit light-field induced modulations such as the dynamical Franz-Keldysh~\cite{Lucchini2016a,Otobe2016} and Stark~\cite{Sie2015} effects among others. Although not considered here as the pump and probe pulses used do not overlap in time, these types of nonlinear optical phenomena are in principle directly accessible from both KS and GKS RT-TDDFT subject as always to the accuracy limits of specific XC approximations employed.

\section{Conclusions}
In summary, this work investigated the use of range-separated hybrid functionals employing both short- and long-range corrected (SLRC)  nonlocal screened exchange contributions in the context of real-time TDDFT simulations of solid state systems. It was shown that in a prototypical strongly excitonic system like 2D h-BN, the use of SLRC potentials within a generalized Kohn Sham TDDFT framework can enable a balanced treatment of both valence and core excitations simultaneously improving absolute spectral energy positions as well as overall lineshapes relative to semi-local XC approximations. The use of such SLRC potentials to directly simulate attosecond transient absorption spectra of core-excitations was illustrated. Following laser driven electron-hole excitations at energies near the first valence excitonic peak of 2D h-BN, nonequilibrium absorption spectra at the B and N K-edges were shown to exhibit characteristic features consistent with population induced state blocking and transient core-level energy shifts. The findings in this work suggest that the range-separated hybrid functional form in spite of certain limitations has sufficient flexibility to provide a practical and computationally efficient methodology for improving the accuracy of TDDFT based time-domain spectroscopy simulations in condensed matter.  

\section*{Acknowledgments}
This work was supported by the U.S. Department of Energy, Office of Basic Energy Sciences, Division of Materials Sciences and Engineering, under Contract No. DE-AC02- 76SF00515 through TIMES at SLAC. This research used resources of the National Energy Research Scientific Computing Center (NERSC), a U.S. Department of Energy Office of Science User Facility operated under Contract No. DE-AC02-05CH11231.

\section*{References}
\bibliography{lcoe.bib}

\end{document}